\documentclass[12pt,twoside]{article}   
\usepackage{latexsym} 
\usepackage{mathptm}
\usepackage{amsmath}
\usepackage{amssymb}
\usepackage{graphicx}

\begin{document}
 
\title{At the Frontier of Knowledge}
\author{Sabine Hossenfelder \thanks{hossi@nordita.org}\\
{\footnotesize{\sl Nordita, Roslagstullsbacken 23, 106 91 Stockholm, Sweden}}}
\date{}
\maketitle
 

At any time, there are areas of science where we are standing at the frontier of knowledge, and can wonder whether we have reached a fundamental limit to human understanding. What is ultimately possible in physics? I will argue here that it is ultimately impossible to answer this question. For this, I will first distinguish three different reasons why the possibility of progress is doubted and offer examples for these cases. Based on this, one can then identify three reasons for why progress might indeed be impossible, and finally conclude that it is impossible to decide which case we are facing. 

\section*{Doubt}

There are three different reasons why scientists question whether progress in a particular direction is possible at all. 

\begin{itemize}
\item[D1)] {\bf There exists a proof or no-go theorem for theoretical impossibility.}

Modern theoretical physics is formulated in the language of mathematics, and consequently subject to mathematical proof. Such proof can be in the form of excluding particular scenarios due to inconsistency. 

A basic example is that in Special Relativity it is not possible for massive particles to travel faster than the speed of light. Other examples are the Weinberg-Witten theorem that shows the incompatibility of massless gravitons with any Lorentz covariant renormalizable quantum field theory and with that constrains approaches to emergent gravity \cite{Weinberg:1980kq}, or the no-go theorems on gravitational theories with more than one interacting metric tensor \cite{Boulanger:2000rq}. 

Physicists have a love-hate relationship with no-go theorems. They love them for the power to sort out possible options and reduce the space of theories that have to be considered. No-go theorems also clarify why a particular direction is not promising. Physicists hate no-go theorems for the same reasons.

\item[D2)] {\bf There exists an argument for practical impossibility.}

Even though progress may not be impossible on theoretical grounds, it may be impossible for all practical purposes. An example may be testing quantum gravity.  To present day we have no experimental evidence for quantum gravity. And as if that wasn't depressing enough already, it has been shown \cite{grav} that even with a detector the size of Jupiter we would not be able to measure a single graviton if we waited the lifetime of the universe, and any improvement in the detector would let it collapse to a black hole. It is of little comfort that we could test particle scattering in the regime where quantum gravitational effects are expected to become important with a collider the size of the Milky Way. 

Another example for questioning practical possibility is the emergence of structures on increasingly macroscopic levels. While most particle physicists believe in reductionism and would insist the atomic structures, molecule properties, and chemical reactions can in principle all be derived from the Standard Model of Particle Physics, we are far off achieving such a derivation.  Even more glaring gaps arise on higher levels. Can one derive all of biology from fundamental physics? What about psychology? Sociology, anybody? A hardcore believer in reductionism will think it possible.

It has been shown in a specific setting that more really {\sl is} different \cite{different} and
a derivation of emergent from fundamental properties impossible even theoretically. This specific setting is an infinitely extended, and thus unphysical, system but nevertheless sharpens the question for practical possibility even for finite systems. This example is still under debate, but it might turn out to also represent a case in which for all practical purposes a derivation is impossible to achieve. 

\item[D3)] {\bf Despite long efforts, no progress has been made.}
 
This situation is one that seems to bother physicists today more than ever due to the lack of breakthroughs in fundamental physics that has lasted several decades now. This is even more frustrating since meanwhile the world around us seems to change in a faster pace every day.

As an example for doubt of this category may serve the understanding of quantum mechanics, in particular its measurement process and interpretation. ``Shut up and calculate'' is a still prevalent pragmatic approach frequently complemented by the attitude that there is nothing more to understand than what our present theories describe, and all questioning of the foundations of quantum mechanics is eventually nothing but a waste of time, or a pastime for philosophers, or maybe both amounts to the same. 

Another example is instead of attempting to explain the parameters of the Standard Model to conjecture there is no explanation other than that we just happen to live in a part of the ``multiverse'' -- a structure containing universes with all possible choices of parameters -- in which the parameters are suitable for the existence of life. After all, if life wasn't possible with the parameters we observe, then we wouldn't be here to observe them. While this is an expression of doubt of category 3, it is not to say invoking such reasoning, known as the ``anthropic principle,'' is necessarily scientifically empty. We have observational evidence that our universe allows for the existence of life, and given a useful quantification of ``existence of life,'' the requirement of its possibility constrains the parameters in the standard model. The controversy remains though whether or not to give up searching for a more satisfactory explanation \cite{Smolin:2004yv} simply on the basis that none has been found for many decades. 

\end{itemize}

\section*{Impossibilities}

The previous section categorized causes to suspect fundamental limits to our knowledge; the following categorizes actual reasons for impossibilities corresponding to the above mentioned three reasons of doubt.

\begin{itemize}
\item[I1)] {\bf Impossible because the laws of Nature do not allow it.}

That is D1 is indeed true. This implies D2 is also true.

\item[I2)] {\bf Possible in theory, but impossible in practice.}

That is though D1 is not true, D2 is true.

\item[I3)] {\bf Possible both in theory and in practice, but not yet possible}

Though progress is not excluded neither in theory nor practice, it might not be possible at a given time because theoretical knowledge 
is still missing, or necessary data has not yet been obtained. 
Scientific insight builds upon previous knowledge. Progress is thus gradual and can stagnate if an essential building block is missing.
\end{itemize}

Since impossibility of the type I3 can be overcome, we will not consider it to be a fundamental impossibility.

Though physicists do not usually include this in discussions about fundamental limits, any question for what is possible should take into account constraints set by the human brain. It is in our nature to overestimate the human capability to understand the world and to exert control about it. However, the capacity and ability of our brains is finite. It is limited in the processes it can perform, and it is limited in speed. There will thus be problems the human brain in its present form will not be able to solve at all, or not in a limited amount of time. And since solving a problem might be necessary to sustain an environment in which solving of problems can be pursued, a problem that cannot be solved in a limited amount of time might turn into one that cannot be solved at all.

This limit could be overcome with improvements of the human brain, either by evolution or design. It is far from clear though whether such an improvement can be limitless. Though not in the realm of physics, the possibility of enhancements of human cognition is another question at the frontier of knowledge to which we presently have no answer. The above cases I2 and I3 both should be understood as including this potential limit to human ingenuity: An experiment that we cannot think of cannot be realized, and a problem whose solution takes more time than has passed will not yet be solved.

\section*{Where are we?}

Let us now investigate whether from any of the three doubts one can conclude a fundamental impossibility of type I1 or I2. 

First, we note that proofs are only about the mathematical properties of certain objects in
their assumptions. A physical theory that describes the real world necessarily also needs a connection between these mathematical objects and the corresponding objects of the real world. While evidence might be abundant that a particular mathematical description of reality is excellent, it can never 
be verified and shown to be true. Consequently, it is impossible to know whether a particular mathematical representation is indeed a
true description of reality, and it cannot be concluded a mathematical proof based on it must necessarily apply to the physical world. We can thus never know whether D1 is caused by an actual fundamental impossibility I1. 

Another way to put this is that no proof is ever better than its assumptions. A loophole in the Coleman-Mandula theorem \cite{Coleman:1967ad} feeds today a huge paper-producing industry called Supersymmetry, and bi-metric theories may be non-interacting \cite{Hossenfelder:2009ne}. 

Turning to doubt D2, no argument for practical impossibility can be obtained without a theoretical basis quantifying this practicability. Since the theoretical basis can never be verified, neither can the practical impossibility. Thus,
I2 cannot be followed from D2, and neither of the both fundamental impossibilities can ever be identified with certainty. 

Coming back to our earlier D2 examples, despite all ridicule about ``Chaoplexity'' \cite{plex} scientists still investigate 
emergence in complex systems with the hope to achieve a coherent understanding, and during the last decade an increasing amount of tests of quantum gravity has been proposed. These proposals have in common a modification in the assumptions that lead to the conclusion quantum gravity might be practically untestable. Two scenarios that have obtained particularly much attention are higher dimensional gravity, in which case
quantum gravity might become testable at the Large Hadron Collider \cite{Landsberg:2008ax}, and deviations from Lorentz invariance resulting in modified dispersion relations \cite{AmelinoCamelia:2008qg}, potentially observable in gamma ray bursts \cite{AmelinoCamelia:2009pg}. Depending on your attitude you might call these studies interesting or a folly, but what they are for certain is possibilities.

Finally, let us discuss doubt D3. If we assume knowledge discovery is pursued as a desirable activity then doubt D3 is equivalent to impossibility I3, since in this case the only reason for lacking progress can be that it has not been possible. With hindsight one often wonders why a particular conclusion was not drawn earlier, even though the pieces were all there already. But since we included limitations of the human brain, slow insights represent imperfections in scientists' thought processes that are part of I3. So long as increasing the understanding of Nature continues to be part of our societies' pursuits, it can then never be concluded from D3 that an impossibility is fundamental. 

What we can thus state with certainty at any time is merely ``To our best current knowledge...'' To our best current knowledge it is not possible to travel faster than the speed of light. To our best current knowledge we cannot see beyond the black hole horizon. To our best current knowledge the measurement process in quantum mechanics is non-deterministic. 

It remains to be said however that progress on fundamental questions becomes impossible indeed if we do not pursue it. And one reason for 
not pursuing it would be the mistaken conviction that it is impossible. 

Scientific progress is driven by curiosity, and the desire to contribute a
piece to mankind's increasing body of knowledge. It lives from creativity, from stubbornness, and from hope. What I have shown here is that there is always reason to hope.

\end{document}